\documentclass[twocolumn,showpacs,floatfix,prl,aps]{revtex4}
\usepackage{float}
\usepackage{graphicx}

\begin{document}
\title{Topology, Delocalization via Average Symmetry \\ and the Symplectic Anderson Transition}

\author{Liang Fu}
\affiliation{Dept. of Physics, Massachusetts Institute of Technology,
Cambridge, MA 02139}
\author{C. L. Kane}
\affiliation{Dept. of Physics and Astronomy, University of Pennsylvania,
Philadelphia, PA 19104}

\begin{abstract}

A field theory of the Anderson transition in two dimensional disordered systems with spin-orbit interactions and time-reversal symmetry is developed, in which the proliferation of vortex-like topological defects is essential for localization.  The sign of vortex fugacity determines the $Z_2$ topological class of the localized phase.
There are two distinct, but equivalent transitions between the metallic phase and the two insulating phases.  The critical conductivity and correlation length exponent of these transitions are computed in a $N=1-\epsilon$ expansion in the number of replicas, where for small $\epsilon$ the critical points are perturbatively connected to the Kosterlitz Thouless critical point.   Delocalized states, which arise at the surface of weak topological insulators and topological crystalline insulators, occur because vortex proliferation is forbidden due to the presence of symmetries that are violated by disorder, but are restored by disorder averaging.

\end{abstract}

\maketitle

Topology can have a profound impact on Anderson localization in
disordered electronic systems. This was first understood in the
integer quantum Hall effect\cite{laughlin81,halperin82}, where the two dimensional (2D) bulk states at the plateau transition are extended, even in the presence of strong disorder.
In the field theory of localization, this delocalization is associated with the presence of
a topological $\theta$ term in the nonlinear sigma model (NL$\sigma$M)\cite{pruisken}.
At long distances, $\theta$ generically renormalizes to
integer multiples of $2\pi$ reflecting the topological quantization of the Hall conductivity, while
$\theta=\pi$ is a critical point, at which bulk states are extended.  More recently, this paradigm for delocalization has been extended to other symmetry classes and dimensions\cite{fendley,schnyder08}.

It is now known that there are two topologically distinct classes of 2D
insulators in the presence of time reversal (TR) symmetry\cite{km,hk,qz}.
A field theory of localization in the symplectic class should incorporate and distinguish the two insulating states.
Such a theory must go beyond the single-parameter scaling based on conductivity alone\cite{g4, wegner79, efetov80, hikami81}, but it has been unclear what additional parameter plays the role of $\theta$.

A related difficulty is revealed by recent studies of surface states of 3D
{\it weak} topological insulators (WTI)\cite{ringel,mong,liu12} and of topological
crystalline insulators (TCI)\cite{fu11,hsieh12}.  These surfaces have an {\it even} number of 2D Dirac fermions, so the $Z_2$ topological term that guarantees delocalization of a 3D strong topological insulator (STI) surface\cite{ryu07,schnyder08} is absent.
Nonetheless, general
arguments, as well as numerics\cite{mong}, suggest that contrary to the conventional picture, the surface remains delocalized even with strong disorder, due to the
existence of discrete symmetries that are violated by disorder, but remain
unbroken on average.
This poses the question of how average symmetries fit into the
field theory of localization, and how those symmetries prevent localization.

In this paper we answer those questions by examining the crucial
and unexplored role played by topological defects in the NL$\sigma$M in the 2D symplectic class\cite{mirlin}.  We show that
the proliferation of pointlike
vortices is essential for localization, and that the sign in the partition function associated
with those vortices distinguishes a topological insulator (TI) from a trivial insulator in 2D.
The ensemble average symmetry forces the system to be on a line where the fugacity of vortices vanishes and hence dictates delocalization.
In addition to explaining the origin of the metallic state, this analysis provides new insight into the 2D symplectic metal-insulator transition.  We show that there are two distinct but equivalent fixed points that describe transitions to insulator and TI states.  Moreover, by treating the number of replicas, $N$ as a continuous variable, we show that for $N=1-\epsilon$ the Anderson transition fixed points are perturbatively connected to the Kosterlitz-Thouless (KT) transition fixed point\cite{kosterlitz} for $\epsilon\rightarrow 0$.  This allows us to compute the critical conductivity and correlation length exponent perturbatively in an $\epsilon$ expansion.

Before describing the symplectic class, we briefly discuss a simpler version of delocalization via average symmetry in the unitary class.  It is well known that the surface of a TR invariant 3D STI is delocalized\cite{ryu07,fk07}, but TR violating perturbations can lead to localization.   This localized state is in a sense ``half" of a quantum Hall state, and has $\sigma_{xy} = \pm e^2/2h$.  Importantly, the time reverse of this state, with $\sigma_{xy} = \mp e^2/2h$, is topologically distinct.  If impurities have random local moments, TR symmetry remains unbroken on average. This places the system precisely at the transition between the two localized states.
This can be modeled by considering an ensemble in which each member violates TR, but the whole ensemble is TR invariant.  Such a system is described by a NL$\sigma$M in the unitary class, which in 2D allows a Pruisken $\theta$-term.
Since TR, when applied to the ensemble, takes $\theta$ to $-\theta$, average TR symmetry constrains $\theta$ to be $0$ or $\pi$.  The surface of a STI corresponds to $\theta=\pi$, so the system flows to the delocalized critical point associated with the plateau transition.  If the average symmetry is broken by an applied magnetic field, then the system flows to a localized phase with $\sigma_{xy} = \pm e^2/2h$. It would be interesting to probe this experimentally, and to compare the conductivity and critical behavior as a function of applied magnetic field with integer quantum Hall transitions.

The surface of a WTI or a TCI also possesses an average symmetry.   For the WTI, the symmetry is translation by one lattice constant in the direction of the layering of the WTI, while for the TCI studied in Ref. \onlinecite{hsieh12}, it is a mirror symmetry.  By breaking this symmetry, it is possible to gap the surface, leading to localization.  However, applying the discrete symmetry to the gapped state leads to a topologically distinct localized state.  Interfaces between the two localized states are associated with a helical edge mode.  If the discrete symmetry is respected on the average, then the system is precisely on the boundary between the two localized states---analogous to the transition between the 2D TI and trivial insulator.  Unlike the unitary case,
this transition need not occur at a point, but can involve an intermediate metallic phase.   It is clear, however, that even for strong disorder, the system can not be localized at this point because it is impossible to change the topological class without extended states crossing the Fermi energy.

To develop a field theory for this delocalization, we use the fermionic replica theory introduced by Efetov et al.\cite{efetov80}.  Our analysis closely parallels that of Ryu, et al.\cite{ryu07}.   We consider a system with average Hamiltonian ${\cal H}_0$, along with gaussian correlated TR invariant disorder.  Using the replica trick to integrate out the disorder, the disorder averaged product of retarded and advanced Green's functions can be generated from the partition function $Z = \int D[\bar\psi,\psi] e^{-S}$, with
\begin{equation}
S = \int d^2 r[
\bar\psi_a(({\cal H}_0 - E)\delta_{ab} + i\eta \Lambda_{ab})\psi_b - {g\over 2} (\bar\psi_a \psi_b)( \bar\psi_b \psi_a)]
\label{zz}
\end{equation}
Here $a=1, ..., 2N$ is an index for $N$ retarded and $N$ advanced replicas,
and $\Lambda = 1_N \oplus (-1)_N$, where $1_N$ is a $N\times N$ identity matrix.
$\psi_a$ is a Grassman field, which includes (suppressed) spin, position and possibly orbital indices.
$\bar\psi_a \equiv \psi_a^T i\sigma^y $, where $\sigma$ acts
on the spin indices.  TR symmetry requires $\sigma^y {\cal H}_0 \sigma^y = {\cal H}_0^*$, so that $i\sigma^y (H_0-E)$ is a skew symmetric matrix.  For $\eta = 0$, (\ref{zz}) is invariant under $O(2N)$ rotations among the replicas, which is broken down to $O(N) \times O(N)$ by 
$\eta$.

A theory of the Nambu Goldstone modes associated with this symmetry breaking is formulated by Hubbard Stratonovich decoupling the four fermion interaction, and performing a saddle point expansion about the broken symmetry state.  After freezing the massive modes, the saddle point is characterized by a $2N \times 2N$ matrix field $Q = O^T \Lambda O$, with $O \in O(2N)$.  Distinct values of $Q$ belong to the coset
$G/H = O(2N)/O(N)\times O(N)$ and satisfy $Q = Q^T$, $Q^2 = 1$.  A theory for the long wavelength fluctuations in $Q_{ab}$ is obtained by integrating $\psi_a$ in the background of a spatially varying $Q_{ab}$.  This gives
$
Z_{\rm eff} = \int D[Q] e^{-S_{\rm eff}[Q]}
$ with
\begin{equation}
e^{-S_{\rm eff}[Q]} = \int D[\bar\psi,\psi] e^{-\int d^2 r[
\bar\psi_a\left[ ({\cal H}_0 - E)\delta_{ab} +  i\Delta Q_{ab}\right]\psi_b]}
\label{seff}
\end{equation}
Here $\Delta$ is a parameter characterizing the bare scattering time that is determined self consistently at the saddle point.  Expanding in gradients gives the NL$\sigma$M,
\begin{equation}
S_{\rm eff}^0[Q] = {1\over {32\pi t}} \int d^2 r{\rm Tr}[ (\nabla Q)^2 ],
\label{nlsm}
\end{equation}
where the coupling constant $t$ characterizes the disorder strength and is related at lowest order to the resistivity, $\sigma = (2\pi t)^{-1} e^2/h$.   The renormalization of $t$ at long wavelengths is described by the perturbative renormalization group (RG) equation\cite{efetov80,hikami81,brezin79,wegner89,tfootnote}
\begin{equation}
dt/d\ell = \beta(t) , \quad  \beta(t) = 2(N-1) t^2 + ...
\label{beta(t)}
\end{equation}
In the replica limit, $N\rightarrow 0$, the weak coupling fixed point $t=0$ is stable, indicating the stability of the symplectic metal phase, characterized by weak antilocalization.

Eq.(\ref{nlsm}) is not the whole story because topologically non trivial configurations of $Q$ can have important non-perturbative effects.
There are two types of topological configurations associated with the first and second homotopy group of $G/H$.
Ryu {\it et al.}\cite{ryu07} argued that $\pi_2(G/H) = Z_2$ allows for a topological term with $\theta=\pi$ associated with the $Z_2$ global anomaly. Such a term
applies to the surface states of 3D STIs, but is absent in 2D systems as well as the WTI or TCI surface because of fermion doubling.

For our problem, the crucial topological objects are point-like defects in 2D similar to vortices which are allowed by the nontrivial $\pi_1(G/H)=Z_2$.  We now argue that these defects are necessary for localization, and that their contribution to $Z_{\rm eff}$ encodes the distinction between a trivial and topological insulator.
The role of vortices can be understood by considering an inhomogeneous 2D system in which a TI in region $S$ with boundary $C$ is surrounded by a trivial insulator, as shown in Fig. 1(a).  Consider the effect of integrating out $\psi_a$ in (\ref{seff}) for weak disorder in the presence of a vortex configuration $Q_{ab}({\bf r})$.
Since the interior of $S$ has a finite band gap, dominant contribution to the action comes from the helical edge states at the boundary $C$.
On $C$, $Q_{ab}({\bf r} \in C)$ is a non-singular and  non-contractible  configuration corresponding to the nontrivial element of $\pi_1(G/H)$.
Repeating the analysis of Ryu {\it et al.}\cite{ryu07} in 1D, one can explicitly verify that the delocalization of the helical edge states is associated with a topological term in the 1D NL$\sigma$M\cite{schnyder08},
\begin{equation}
e^{-S_{eff}[Q]} \propto (-1)^{n(C)}
\end{equation}
where $n(C) = 0,1$ is the $Z_2$ homotopy class of $Q$ on $C$.  Importantly, since $Q$ is defined in all space (except at the cores of vortices), $n(C)$ may be viewed as the number of vortices in $S$ mod 2.  This leads to a {\it bulk} characterization of the TI based on the 2D NL$\sigma$M:   {\it in the TI the fugacity $v$ of $Z_2$ vortices is negative.}  In the trivial insulator, the topological term is absent, and $v$ is positive.  The positivity of $v$ for a trivial insulator can also be understood in the limit of vanishing spin-orbit coupling in $H_0$: the fermion determinant from integrating out $\psi_a$  is squared due to spin degeneracy.

\begin{figure}
\includegraphics[width=3in]{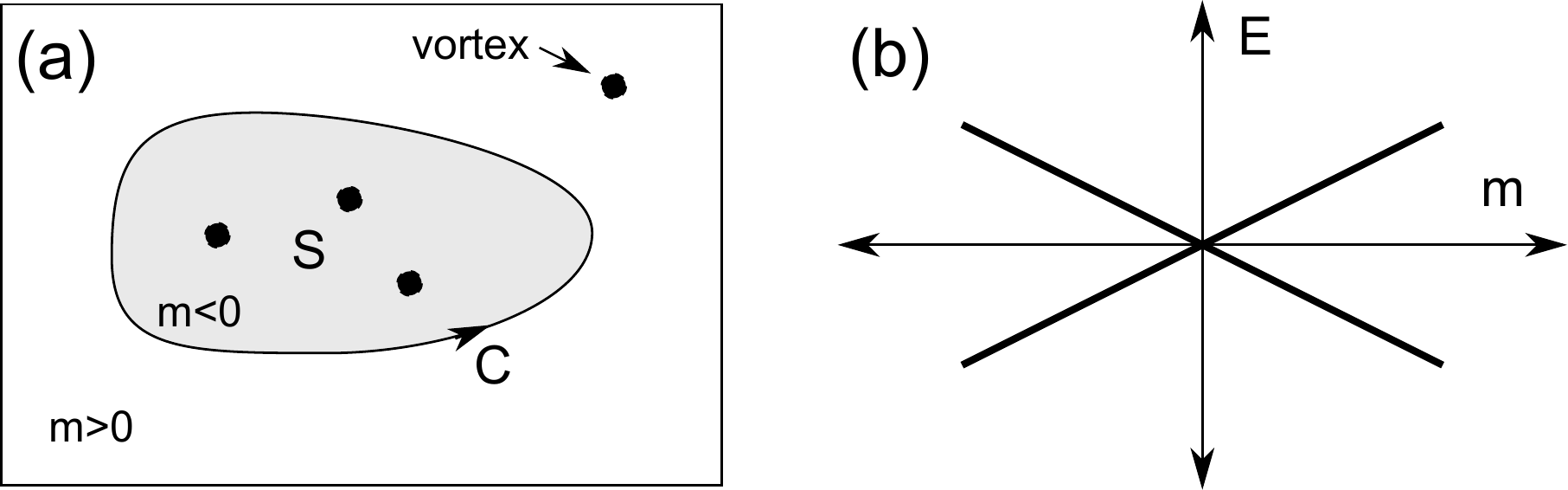}
\caption{(a) A 2D TI ($m<0$) in region S with boundary C is surrounded by trivial insulator ($m>0$).  The sign of the partition function depends on the parity of the number of vortices in $S$.  (b) For $m = 0$ the eigenvalues of ${\cal H}^{\rm eff}$ in (\ref{heff})
exhibit a linear zero crossing, which leads to a vanishing fugacity for vortices.}
\label{Fig1}
\end{figure}

At the transition between the trivial and TI, $v$ must pass through zero.
This suggests that at the WTI surface, $v=0$.  We will now proceed to demonstrate this explicitly.
To model the WTI surface, consider a system of helical edge modes $H = v \sigma_x k_x$ that are stacked in the $y$ direction with separation $a$.  Coupling between neighboring layers gaps the surface, except at two Dirac points at $(k_x,k_y) = (0,0)$ and $(0,\pi/a)$.  If we index the two Dirac points by $\tau_z = \pm 1$, then the surface states are described by the Dirac Hamiltonian,
\begin{equation}
H_0 = v (\sigma_x k_x + \sigma_y \tau_z)  + m \sigma_y \tau_y
\end{equation}
The discrete translation symmetry of the WTI corresponds to translating by $a \hat y$, described by $\exp (i p_y a)  = \tau_z$.  We have included a uniform mass term $m \sigma_y \tau_y$\cite{mong}, that describes a translation symmetry breaking dimerization of the layers.  This is the only mass term that respects TR symmetry.  The two topologically distinct dimerization patterns are distinguished by the sign of $m$.  (Including a large $k$ regularization $\sigma^y \tau^y k^2$, this model also describes the 2D transition between a trivial ($m>0$) and topological ($m<0$) insulator\cite{bhz}.)

In general, the NL$\sigma$M (\ref{nlsm}) should include a sum over vortex configurations in $Q$.  The fugacity of the vortices is determined by comparing (\ref{seff}) in the presence and absence of vortices.  To evaluate (\ref{seff}) for a vortex consider the simplest vortex configuration involving a single retarded and advanced pair of replicas.  This can be expressed in terms of a one parameter family of $Q$'s of the form
\begin{equation}
Q(\theta) = 1_{N-1} \oplus \left(\begin{array}{cc} \cos\theta & \sin\theta \\ \sin\theta & -\cos\theta \end{array}\right) \oplus 1_{N-1}.
\label{q(theta)}
\end{equation}
A $Z_2$ vortex is then a configuration where $\theta$ winds by an odd multiple of $2\pi$.

The Grassman integral in (\ref{seff}) defines a Pfaffian, so that the vortex fugacity may be written
\begin{equation}
v = { {\rm Pf}[i \sigma^y D(Q)] \over{ {\rm Pf}[i \sigma^y D(Q_0)]}},
\end{equation}
Where $Q$ is a vortex configuration, and $Q_0=\Lambda$.
In the space of the two nontrivial replicas we have
\begin{equation}
D(Q) = ({\cal H}_0 - E) + i\Delta ( \mu^z \cos\theta + \mu^x \sin\theta ).
\end{equation}
Here $\mu_z$ is a Pauli matrix in the space of the two nontrivial replicas.
To evaluate the Pfaffian, we use a trick similar to that used by Ryu {\it et al.}\cite{ryu07}, and compute $({\rm Pf}[i\sigma^y D])^2 = {\rm det}[i\sigma^y D] = {\rm det}[\mu^y D]$.  This is useful because $\mu^y D \equiv {\cal H}^{\rm eff}$ is a Hermitian operator given by
\begin{equation}
{\cal H}^{\rm eff} = \mu^y({\cal H}_0 - E) + \Delta( \mu^x \cos\theta - \mu^z \sin\theta),
\label{heff}
\end{equation}
so the determinant is the product of its real eigenvalues.  The TR symmetry of the original ${\cal H}_0$ becomes a particle-hole symmetry,  $\{ {\cal H}^{\rm eff},\Xi\} = 0$, with $\Xi = \mu^y \sigma^y K$.  Moreover, when $m=0$, ${\cal H}^{\rm eff}$ decouples into two independent Hamiltonians for $\tau^z =\pm 1$.  Each is identical to a topological superconductor in class D, with $\theta$ playing the role of the superconducting phase.  It follows that there are two zero modes indexed by $\tau^z = \pm 1$ bound to the core of a vortex with odd-vorticity.  For $m\ne 0$, the zero modes are coupled and split, as shown in Fig. 1b.  We conclude that ${\rm det}[\mu^y D]$ has a second order zero at $m=0$, so ${\rm Pf}[i\sigma^y D]$ has a first order zero, which involves a sign change as a function of $m$.
This shows that sign of the fugacity $v$ of the vortices depends on $m$ and vanishes for $m=0$.
Thus isolated $Z_2$ vortices are {\it forbidden} at the surface of a WTI.   More generally, it is possible to have multiple vortices, since in that case the zero modes will split even for $m=0$, leading to a nonzero Pfaffian.  However, since the splitting vanishes exponentially in the separation, the vortices will be confined by a linear potential.

It is thus clear that the vortex fugacity $v$ is a crucial variable that must be included in the NL$\sigma$M.  The topological and trivial insulators are distinguished by the sign of $v$. Proliferation of vortices leads to a disordered state in the corresponding topological class.  For $v=0$, qualitatively different behavior  is expected to occur reflecting the delocalization of  the WTI or TCI surface states.
In this case, the target space of the NL$\sigma$M effectively lifts to its double cover,
$\tilde G/\tilde H = SO(2N)/SO(N)\times SO(N)$, for which $\pi_1(\tilde G/\tilde H)=0$.  Since $G/H$ and $\tilde G/\tilde H$ have identical local structure, their perturbative $\beta$ functions will be identical.  It is useful to consider this behavior as a function of the replica number, $N$.

For $N>1$, $\beta(t) >0$, and the weak coupling fixed point is unstable, leading to a disordered phase even in the absence of vortices.  This phase is ``less disordered" than the disordered phase with $v \ne 0$, though.  The confinement of $Z_2$ vortices leads to a topological order similar to a $Z_2$ spin liquid\cite{rokhsar}.
This can be clearly seen by placing the system in a torus: there are four topologically disconnected sectors corresponding to the homotopy classes of $Q \in G/H$ along the two large loops.
When $v$ is turned on in this disordered phase, the $Z_2$ vortices immediately condense.  The $v=0$ line thus describes a first order transition between the $v>0$ and $v<0$ phases.

The behavior for $N\rightarrow 0$ is expected to be qualitatively different.  In this case the weak coupling fixed point is stable, and we know from the arguments presented above that even at strong coupling there must be delocalization for $v=0$.  It is useful to consider the critical value $N=1$ that separates these behaviors.   The theory for $N=1$ is simply the XY model, and
$Q$ is fully parameterized by $\theta$ in (\ref{q(theta)}).  The action (\ref{nlsm}) becomes
\begin{equation}
S_{N=1} = {1\over {16\pi t}} \int d^2 r (\nabla\theta)^2.
\end{equation}
Since the target space, $S^1$, is flat, $\beta(t)=0$ to all orders, but vortices
modify the behavior.
For small $t$, $2\pi$ vortices in $\theta$ are bound, and the system flows to a fixed line parameterized by $t$.  For $t > t^* = 1/16$ vortices unbind at a KT transition\cite{kosterlitz} to a disordered phase.

We now consider the behavior for $N<1$, treating $N$ as a continuous variable.  Since $Z_2$ vortices are present for all $N$, it is reasonable to examine their effects as a function on $N$.  We find that the theory can be controlled for $N = 1-\epsilon$, with $\epsilon\ll 1$.  To lowest order in $\epsilon$, the KT flow equations to lowest order in $v$ are modified by the nonzero (but small) $\beta(t) \equiv (N-1) \tilde\beta(t)$,
\begin{eqnarray}
dt/d\ell &=& -\epsilon \tilde\beta(t) + v^2 \nonumber\\
dv/d\ell &=& (2 - (8t)^{-1}) v.
\label{rgflow}
\end{eqnarray}
To this order, we are free to set the coefficient of $v^2$ to one by rescaling $v$.  The RG flows are shown in Fig. 2.  There are two fixed points at
\begin{equation}
t^* = 1/16, \quad\quad v^* = \pm [\epsilon \tilde\beta(t^*)]^{1/2}
\end{equation}
For small $\epsilon$, these fixed points are within perturbative range of the KT fixed point.  These fixed points describe a transition between the ordered and disordered phases of the $O(2N)/O(N) \times O(N)$ NL$\sigma$M for $N < 1$.   For $N\rightarrow 0$ we identify these fixed points with the Anderson transition between the symplectic metal and the localized trivial/topological insulator. These two metal-insulator transitions have identical bulk critical behaviors, which is expected from the fact that the total number of $Z_2$ vortices in a closed system is always even.

\begin{figure}
\includegraphics[width=3in]{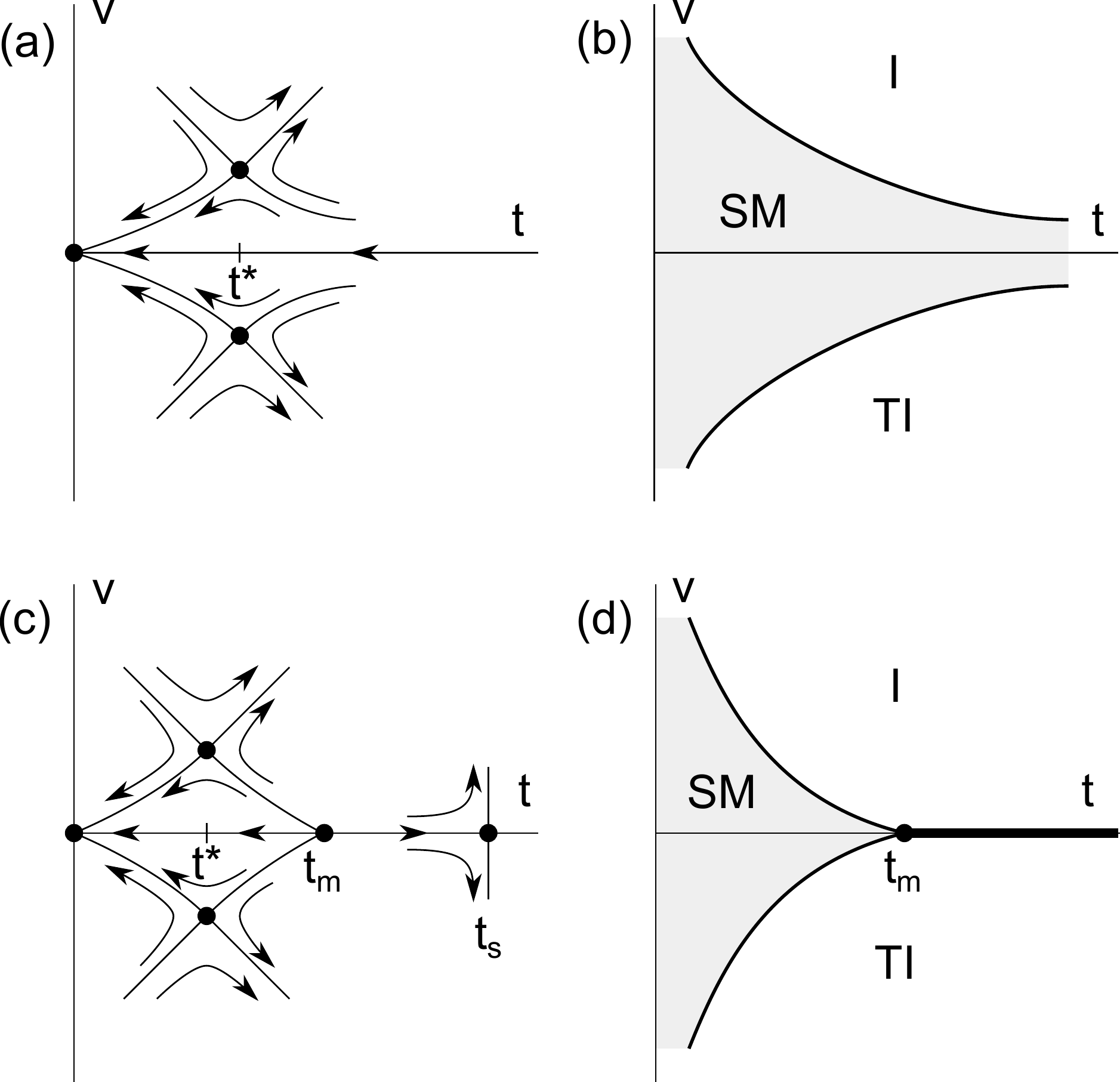}
\caption{(a) RG flow diagram based on (\ref{rgflow}).  The stable fixed point at $(t,v)=(0,0)$ is
the symplectic metal (SM).  The two unstable fixed points at
$(t^*,\pm v^*)$ approach the KT transition for $\epsilon = 1-N \ll 1$ and for $N\rightarrow 0$ are
identified with the Anderson transition. (c) is a flow diagram which includes an third fixed point at $(t_m,0)$, along with a fixed point at $(t_s,0)$ which describes a direct transition between TI and I.  (b) and (d) are phase diagrams corresponding to (a) and (c).}
\label{Fig2}
\end{figure}

By expanding (\ref{rgflow}) about the fixed point, we can identify the critical conductivity and the correlation length exponent associated with the symplectic Anderson transition.  To lowest order in $\epsilon$ we find $\sigma^* = (2\pi t^*)^{-1} e^2/h = (8/\pi) e^2/h$, and $\nu = 2 t^*/(\epsilon \tilde\beta(t^*))^{1/2}$.  While $\tilde\beta(t^*)$ is not known exactly, $\beta(t)$ has been computed perturbatively up to order $t^5$\cite{wegner89}.  The small value of $t^*$ is well within the range of this perturbation theory.  The second order term gives only 6\% correction and the higher terms are even smaller.  Using the first term from Eq. \ref{beta(t)} we find $\nu = (2/\epsilon)^{1/2}$.
Extrapolating to $\epsilon=1$ gives
\begin{equation}
\sigma^* \sim 2.5 e^2/h,  \quad\quad
\nu \sim 1.4.
\end{equation}

These values are rather different from numerical estimates of critical exponents in previous model studies, which give $\sigma^* \sim 1.4 e^2/h$ and $\nu \sim 2.7$\cite{asada,markos,obuse}.
We suggest two possible origins of the discrepancy, depending on the behavior of the $N=0$ NL$\sigma$M at strong coupling, which cannot be accessed in the present analysis.
One possibility is that for $N\rightarrow 0$, $\beta(t)<0$ for all $t$ along the line $v=0$. The corresponding RG flow and phase diagrams are shown in Fig.2a-b.
In this case, the symplectic metal-insulator transition is governed by the fixed point $(t^*, v^*)$. The discrepancy in exponents is
then most likely due to the slow convergence of the $\epsilon$ expansion similar to the $d=2+\epsilon$ expansion for the 3D Anderson transition.

A more interesting possibility is that for $N\rightarrow 0$, $\beta(t)$ changes sign at a critical point $t_m$ on the line $v=0$.
In fact, $t_m$ is present for $N=1-\epsilon$.  For $N=1$, double vortices are allowed, and will in general have non zero fugacity.  The theory with both single and double vortices can be analyzed using a dual sine-gordon theory,
\begin{equation}
S = \int d^2 r {t\over{\pi}}(\nabla\varphi)^2 + v \cos\varphi + v_2 \cos 2\varphi
\end{equation}
where $v_2$ is the fugacity for double vortices.
When $v=0$, $v_2$ becomes relevant at $t_m=1/4$.
When $v_2$ flows to strong coupling, $v=0$ describes a first order transition similar to the case when $N>1$.  However, it is unlikely that this first order transition persists to $N=0$, which is a theory of disordered non-interacting electrons.
Instead, the most likely scenario is that there is a continuous direct transition between trivial and topological insulators, controlled by
a strong coupling fixed point $t_s$, as indicated in Fig. 2c-d. If this is the case, the  flow trajectory near $v=0$ line may be strongly influenced by the two nearby fixed points $t_m$ and $t_s$. Nonetheless, the true scaling behavior of metal-insulator transition in the symplectic class is ultimately controlled by the fixed points $(t^*, v^*)$. In this regard, we note that our correlation length exponent is close to the $\nu=1.6$ found by Onoda et al. \cite{onoda} in a model of a metal to TI transition.

\acknowledgments

We thank Anton Akhmerov, Jens Bardarson and Ady Stern for interesting discussions.  C.L.K was supported by NSF grant DMR 0906175. L.F. was supported by startup funds from MIT.


\begin{thebibliography}{10}
\bibitem{tfootnote}
Our normalization of $t$ in (\ref{nlsm}) and (\ref{beta(t)}) is chosen to be consistent with $\beta(t)$ in Refs. \onlinecite{hikami81, brezin79,wegner89}.  In Ref. \onlinecite{efetov80}, $t$ differs by a factor of $32\pi$, and is rescaled by $2\pi$ in $\beta(t)$.  $N$ differs by a factor of $2$.  The relation between $t$ and $\sigma$ is fixed by comparing (\ref{beta(t)}) with the weak antilocalization correction to $\sigma$.
\end{thebibliography}
\end{document}